\documentclass[fleqn,usenatbib,useAMS]{mnras}

\usepackage{newtxtext}

\usepackage[T1]{fontenc}
\usepackage{ae,aecompl}
\usepackage{graphicx}	
\usepackage{amsmath}	
\usepackage{amssymb}	
\usepackage{multirow}
\usepackage{booktabs}
\usepackage{verbatim}
\usepackage{caption}
\usepackage{adjustbox}
\usepackage{hhline}
\usepackage{float,caption}
\usepackage{natbib}
\usepackage{titlesec}
\usepackage{subcaption}	
\usepackage[nameinlink,noabbrev]{cleveref}
\usepackage[fleqn]{nccmath}

\title[]{Non-Gaussianity Constraints using Future Radio Continuum Surveys and the Multi-Tracer Technique}

\author[Z. Gomes et al.]{Zahra Gomes,$^{1}$\thanks{E-mail: zahra.gomes@physics.ox.ac.uk}
Stefano Camera,$^{2,3,4,5}$
Matt J. Jarvis,$^{1,5}$
Catherine Hale$^{1,6}$
\and
and Jos\'e Fonseca$^{7,8,5}$
\\
$^{1}$Oxford Astrophysics, Department of Physics, Keble Road, Oxford, OX1 3RH, UK\\
$^{2}$Dipartimento di Fisica, Universit\`a degli Studi di Torino, via P. Giuria 1, 10125 Torino, Italy\\
$^{3}$INFN -- Istituto Nazionale di Fisica Nucleare, Sezione di Torino, via P. Giuria 1, 10125 Torino, Italy\\
$^{4}$INAF -- Istituto Nazionale di Astrofisica, Osservatorio Astrofisico di Torino, strada Osservatorio 20, 10025 Pino Torinese, Italy\\
$^{5}$Department of Physics, University of the Western Cape, Bellville 7535, South Africa\\
$^{6}$CSIRO Astronomy and Space Science, 26 Dick Perry Avenue, Kensington, Perth, 6151, WA, Australia\\
$^{7}$Dipartimento di Fisica e Astronomia ``G. Galilei'', Universit\`a degli Studi di Padova, via Marzolo 8, 35131 Padova, Italy\\
$^{8}$INFN -- Istituto Nazionale di Fisica Nucleare, Sezione di Padova, via Marzolo 8, 35131, Padova, Italy 
}

\date{Accepted XXX. Received YYY; in original form ZZZ}

\pubyear{2018}

\begin{document}
\label{firstpage}
\pagerange{\pageref{firstpage}--\pageref{lastpage}}
\maketitle

\begin{abstract}
Tighter constraints on measurements of primordial non-Gaussianity will allow the differentiation of inflationary scenarios. The cosmic microwave background bispectrum---the standard method of measuring the local non-Gaussianity---is limited by cosmic variance. Therefore, it is sensible to investigate measurements of non-Gaussianity using the large-scale structure. This can be done by investigating the effects of non-Gaussianity on the power spectrum on large scales. In this study we forecast the constraints on the local primordial non-Gaussianity parameter $f_{\rm NL}$ that can be obtained with future radio surveys. We utilize the multi-tracer method which reduces the effect of cosmic variance and takes advantage of the multiple radio galaxy populations which are differently biased tracers of the same underlying dark matter distribution. Improvements on previous work include the use of observational bias and halo mass estimates, updated simulations and realistic photometric redshift expectations, thus producing more realistic forecasts. Combinations of SKA simulations and radio observations were used as well as different redshift ranges and redshift bin sizes. It was found that in the most realistic case the 1 - $\sigma$ error on $f_{\rm NL}$ falls within the range 4.07 and 6.58, rivalling the tightest constraints currently available. 
\end{abstract}


\section{Introduction}
Primordial non-Gaussianity (PNG) describes deviations from a Gaussian random field in the initial density field present after inflation. These small fluctuations are believed to seed the formation of all the large-scale structure present in the universe today as well as the cosmic microwave background (CMB) temperature anisotropies that we observe and in the inflationary scenario they were formed from quantum fluctuations during or just after inflation. Currently, there exists a large  number of models for inflation which all fit with current observations, but because different models provide different predictions for the extent of primordial non-Gaussianity present, this measurement is a valuable tool for discriminating between these models (see \citealt{Bartolo2004Inf} for a review). Most inflationary models predict that the non-Gaussianity depends only on the local value of the potential, such PNG is said to be of the `local type' and it is parameterized by the $f_{\rm NL}$ parameter which quantifies the deviation from a Gaussian random field $\phi$:
\begin{ceqn} 
\begin{align}
 \Phi = \phi + f_{\rm NL} (\phi^2 - \langle \phi \rangle^2).\label{eq:fnl}
\end{align}
\end{ceqn}

Simple slow-roll, single-field inflationary models predict almost Gaussian density fields with $f_{\rm NL} \ll 1$ $(\mathcal{O}(10^{-2}))$ (\citealt{Maldacena2003,Creminelli2004}) while multi-field models allow larger deviations from Gaussianity with $f_{\rm NL} \gtrsim 1$ (\citealt{Lyth2003, Zaldarriaga2004}). The current tightest constraint on the local PNG parameter is $f_{\rm NL} = 0.8\pm 5$ which was obtained using Planck 2015 CMB data (\citealt{Planck2016PNG}). Planck 2018 data (\citealt{Planck2018PNG}) produced the slightly less constraining result $f_{\rm NL} = 0.9\pm 5.1$. This recent result was less constraining due to the use of more realistic polarization simulations with higher noise levels for estimating the errors. In addition, for large-scale structure measurements, general relativistic projection effects and non-linear evolution are expected to produce scale-dependent effects on large-scale structure measurements very similar to those of PNG with $f_{\rm NL}$ of order unity (\citealt{Yoo2010,Jeong2012,Bruni2012,Camera2015GR}). Therefore, in order to start distinguishing between single-field and multi-field inflation, $\sigma (f_{\rm NL})\sim 1 $ is required, and therefore, this is normally the target of such analyses (\citealt{dePutter2017b}).

Local PNG has\textemdash to date\textemdash been measured most precisely using the bispectrum of the CMB temperature anisotropy maps (\citealt{Komatsu2003WMAP,Planck2016PNG,Planck2018PNG}). The main disadvantage of this method are that such measurements are limited by cosmic variance on large scales and by Silk damping on small scales (\citealt{Alvarez2014,Ferraro2015}). Large-scale structure measurements are a viable alternative to the CMB bispectrum as the non-Gaussianity also leaves an imprint on the large-scale structure (LSS). In particular, $f_{\rm NL} > 0$ leads to an increase in the 3D power spectrum on large scales, which corresponds to an increase in the total halo bias on large scales (\citealt{Dalal2008LSS,Matarrese2008LSS,Carbone2008LSS}). Large galaxy surveys (both optical and radio) such as the Sloan Digital Sky Survey (SDSS) and the National Radio Astronomy Observatory (NRAO) Very Large Array (VLA) Sky Survey (NVSS) have been used to measure the clustering of dark matter and constrain the effects of PNG (\citealt{Slosar2008PNG, Bernardis2010LRG, Xia2010NVSS,Ross2013BOSS}). One approach to this LSS measurement is to use quasar populations as these are highly biased dark matter tracers and cover large volumes of space. \citet{Leistedt2014} and \citet{Castorina2019} study the clustering of the SDSS quasar sample and provide the tightest constraints with LSS methods to date. Combinations of galaxy surveys and CMB maps have also been considered (\citealt{Giannantonio2014}). 

As this PNG effect is most prominent on large scales, very large surveys are used, but these are still limited by cosmic variance on the larger scales and forecasts of the tightest constraints possible with single tracer surveys have found that $\sigma (f_{\rm NL})\sim 1 $ is not possible. The multi-tracer method, first proposed by \citet{Seljak2009MT} significantly reduces the effect of cosmic variance by cross-correlating multiple dark matter tracers, with different biases located in the same sky volume. These populations have the same underlying density distributions and thus the ratio of their biases will cancel the effects of cosmic variance resulting in a measurement that is only limited by shot noise. This multi-tracer technique, applied to the large upcoming galaxy surveys with multiple differently biased galaxy populations is expected to rival the current CMB constraints (see \citealt{Ferramacho2014,Yamauchi2014MT, Alonso2015redblue}). 

An alternative avenue for probing the effects of PNG on the LSS is intensity mapping and future surveys of this type seem promising due to the combination of their large sky coverage and the potential to push to high redshifts (\citealt{Joudaki2011IM,Camera2013IM,Li2017IM,Fonseca2018}). The multi-tracer method has also been suggested for use on HI intensity mapping surveys. For example, \citet{Fonseca2018} forecast that performing such an analysis on the HI survey from the Square Kilometre Array (\citealt{SKARedBook2018}) and the H$\alpha$ survey from SPHEREx (\citealt{Dore2014}) could lead to constraints on $f_{\rm NL}$ on the order of $\sigma (f_{\rm NL})\sim 1 $. Applying the multi-tracer method to combinations of intensity mapping and photometric galaxy surveys is also possible and was first suggested by \citet{Alonso2015redblue} and \citet{Fonseca2015ZeFon}. More recently the combination of intensity mapping from SKA and galaxy surveys from the Large Synoptic Survey Telescope \citep[LSST;][]{lsstbook} was suggested and investigated by \citet{Witzemann2018} and intensity mapping from MeerKAT and galaxy surveys from the Dark Energy Survey (DES) \citep[DES;][]{dark2005dark} was suggested by \citet{Fonseca2017} and forecasts have found that these analyses should provide constraints more than two times better than Planck (\citealt{Fonseca2017}). Recently a forecast analysis has been performed on the use of the multi-tracer method for combining HI intensity mapping with both photometric galaxy surveys and CMB lensing and the constraints of $\sigma(f_{\rm NL}) \sim 1$ were obtained (\citealt{Ballardini2019}). 

Further research on improving the methods currently being used for measuring the $f_{\rm NL}$ signal is also being done by improving the measurement of the galaxy bispectrum and accounting for systematics (\citealt{Tellarini2016PNG,Welling2016noise,Mueller2017Opt, Karagiannis2018Bispec,Uhlemann2018sphere}). Another example is a recently proposed method which involves the creation of a sample with zero bias, allowing the effect of PNG on the clustering to be measured more easily (\citealt{Castorina2018zerobias}). Similarly, studies on the optimal galaxy survey for providing $f_{\rm NL}$ constraints of $\sigma(f_{\rm NL}) = 1$ are also being done (\citealt{dePutter2017}). 

Radio continuum surveys are appealing for use with multi-tracer techniques for a number of reasons: radio surveys cover very large areas of sky, which is particularly advantageous since the increased bias signal is detected on large scales, and is also useful for reducing shot noise. Just as importantly, radio surveys contain a range of distinct radio galaxy populations that have significant spread in their biases (\citealt{hale2017}). Other advantages of radio continuum surveys are that they allow the observation of galaxies to very high redshifts and without dust obscuration (\citealt{Jarvis2015CSKA}), which has been shown to provide advantages in determining cosmological parameters \citep{Camera2012}. Contrary to intensity mapping, galactic and extragalactic foregrounds are not significant issues for the continuum surveys that use resolved galaxy populations.  On the other hand, one disadvantage of using radio surveys is that it is very difficult to get precise redshifts for the individual sources: either multi-wavelength data or HI 21-cm line measurements are necessary (\citealt{Jarvis2015CSKA}).

The Square Kilometre Array (SKA, \citealt{Carilli2004,Dewdney2009}) will be the world's largest radio telescope with thousands of dishes and antennas distributed over South Africa and Australia. It will consist of a mid-frequency instrument (SKA-MID) and a low-frequency instrument (SKA-LOW) and will span a frequency range of 50 MHz to 15 GHz with unprecedented angular resolution and sensitivity. It will be constructed in two main phases, with the first composed of $10\%$ of the intended eventual collecting area. The SKA's radio continuum survey will cover $2\pi$ steradians and redshifts < 5 while its spectroscopic HI survey will detect galaxies up to redshift $\sim 2$. HI intensity mapping surveys that cover redshifts up to 5 may also be possible with the SKA (\citealt{Santos2015, Quinn2015}).

In this paper we update the work by \citet{Ferramacho2014} who perform a tomographic multi-tracer analysis driven by multiple radio galaxy populations with simulated masses and then use this to forecast the PNG constraints possible with the SKA Phase 1. We will perform a similar analysis, using both the galaxy masses estimated by the $S^3$ simulation (\citealt{Wilman2008SKADS}) and those measured by \citet{hale2017} who use Very Large Array (VLA) COSMOS field observations to estimate the average bias values and corresponding halo masses of the radio galaxy populations. In addition, redshift distributions predicted by both $S^3$ and the more recent Tiered Radio Extra-galactic Continuum Simulation (T-RECS) by \citet{Bonaldi2018TRECS} are used and compared. In section \ref{pngLSS} we present an overview of the theoretical framework used for this analysis, followed by a discussion of our choice of radio galaxy populations in section \ref{galpops}. Our methods are described in section \ref{fisher} and our results and analysis are presented in section \ref{results}. Finally, we conclude in section \ref{conclusions}. We use a fiducial cosmology with $H_0 = 67.74$, $\Omega_\Lambda = 0.6911$, $\Omega_{CDM} = 0.26$, $\Omega_b = 0.05$, $A_s = 2.142$ x $10^{-9}$ and $f_{\rm NL}= 0$ taken from the Planck 2015 results (\citealt{Planck2016PNG}).

\section{Primordial Non-Gaussianity in Large-Scale Structure}\label{pngLSS}

A non-zero local primordial non-Gaussianity signal will increase the power of the halo power spectrum on large scales, corresponding to a scale dependent correction to the halo bias given by the expression below (\citealt{Slosar2008PNG,Dalal2008LSS,Matarrese2008LSS,Carbone2008LSS}):
\begin{ceqn}
\begin{align}
b_h(M,z) = b_L(M,z)+ f_{\rm NL}\delta_c[b_L(M,z)-1] \frac{3 \Omega_mH_0^2}{c^2k^2T(k)D(z)}, \label{eq:halo_bias}
\end{align}
\end{ceqn}
where $b_h$ is the total halo bias, $b_L$ is the Gaussian linear halo bias, $f_{\rm NL}$ is the amplitude of the non-Gaussianity (defined in equation \ref{eq:fnl}), $\delta_c$ is the critical over-density for spherical collapse at redshift $z=0$, $T(k)$ is the linear transfer function and $D(z)$ is the growth function. The primordial non-Gaussianity parameter $f_{\rm NL}$ is therefore normally determined from large-scale structure by measuring the galaxy power spectrum or bispectrum (\citealt{Bernardis2010LRG, Ross2013BOSS}). On the other hand, in situations where spectroscopic redshift data is not available (which will be the case with some current and future large scale surveys with volumes and depths that make it unfeasible to measure spectroscopic redshifts), and only the less accurate photometric redshifts or estimated redshift distributions are present, neither the 3D matter power spectrum nor the bispectrum can be reliably computed. Instead, a tomographic analysis with angular power spectra can be used. The photo-$z$ estimates are used to place the galaxies into redshift bins and then the angular power spectrum, which is the projection of the line-of-sight halos (within a particular redshift bin), into the 2D plane can be computed. Although the angular power spectra do not provide as much information as the full 3D analysis, it is a sufficient estimate. Some analyses that have used this method are: \citet{Slosar2008PNG} and \citet{Xia2010NVSS}. 

For the case of a multi-tracer analysis using these angular power spectra, the statistical information is given by the auto and cross correlation power spectra that are split into multipoles (see \citealt{Huterer2001Cls}):
\begin{ceqn} 
\begin{align}
C_l^{i,j}= \frac{2}{\pi} \int^{k_{\rm max}}_{k_{\rm min}} k^2 P_{\delta}(k)W_l^i(k)W_l^j(k){\rm d}k. \label{eq:Cls}
\end{align}
\end{ceqn}
$P_{\delta}(k)$ is the dark matter power spectrum at redshift 0. $W_l^i$ are window functions that incorporate the redshift distribution (${\rm d}n/{\rm d}z$) and bias $b$ of the tracer $i$ as well as the angular geometry of multipole $l$ via a spherical Bessel function $j_l(kr)$ where $r$ is the comoving radial distance to redshift $z$. The form of these window functions is given below:
\begin{ceqn} 
\begin{align}
W_l^i = \int \frac{{\rm d}n}{{\rm d}z}^i D(z) b^i_h(z)j_l(kr){\rm d}z. \label{eq:window}
\end{align}
\end{ceqn}

In this analysis we will compute these auto and cross power spectra for a variety of tracers, defined by radio galaxy type and redshift bin, using estimated bias functions and redshift distributions given the constraints of the SKA telescope. Following this, we will perform a Fisher analysis of this multi-tracer method to determine the constraints on the $f_{\rm NL}$ parameter that will be possible with the SKA. The relatively large width of the redshift bins used in this analysis allows us to neglect the effects of redshift space distortions.

\section{Choice of Radio Galaxy populations} \label{galpops}

There have been a number of attempts to analyse radio galaxies by population and determine their average dark matter halo masses and biases (\citealt{Wilman2008SKADS,Lindsay2014,Magliocchetti2017clustering,hale2017,Bonaldi2018TRECS}). \citet{Wilman2008SKADS} performed a semi-empirical simulation of the extra-galactic radio-continuum sky covering an area of 20 x 20 deg$^2$ with a flux density limit of 10nJy over a range of frequencies. This simulation was created as part of the SKA Design Study Simulated Skies ($S^3$). Five distinct galaxy populations were identified: star forming galaxies (SFGs), star bursts (SBs), Fanaroff-Riley Class I (FRIs), Fanaroff-Riley Class II (FRIIs) and radio quiet quasars (RQQs) and average halo masses and biases were computed for each population. The very low flux limits of $S^3$ make it easily scalable to the proposed 5 $\mu$Jy flux density limit at 1.4 GHz of SKA Phase 1 and the five distinct populations of radio galaxies make this set-up appealing for forecasting the constraints on $f_{\rm NL}$ from the SKA using the multi-tracer method. This was done by \citet{Ferramacho2014} who used the redshift distributions and the mean masses of these five populations to compute auto and cross correlation angular power spectra and perform a Fisher analysis. 

The Tiered Radio Extra-galactic Continuum Simulation \citep[T-RECS; ][]{Bonaldi2018TRECS} is an updated simulation created for the SKA with a sky area of 25 deg$^2$ and a 150 MHz - 20 GHz frequency range. \citet{Bonaldi2018TRECS} do not provide independent estimates for halo masses of galaxy populations but do compare clustering measurements for the T-RECS sources to \citet{hale2017}, and find that the observed biases for the SFGs are higher than predicted while for the active galactic nuclei (AGN) there is reasonable agreement.

In contrast to the simulated data of $S^3$ and T-RECS, \citet{hale2017} explored the clustering of different radio-selected galaxy populations taken from 3 GHz VLA imaging over the $\sim$2 deg$^2$ COSMOS field (see \citealt{Smolcic2017VLACOSMOSa}) using a 5.5 $\sigma$ cut which corresponds to a mean flux density limit of 22 $\mu$Jy at 1.4 GHz. Using the angular two-point correlation functions of the radio galaxies, they estimated average bias values for the median redshifts of the samples and corresponding halo masses for these galaxy populations. See Figure 10 of \citet{hale2017} for a comparison of the measured biases and the $S^3$ simulated bias trends as a function of redshift. The populations considered were: high and low radio luminosity SFGs, high and low radio luminosity AGN and high and moderate luminosity AGN (HLAGN and MLAGN). The radio galaxies were classified into these categories by \citet{Smolcic2017VLACOSMOSb} who gathered a wealth of multiwavelength (optical, near-UV near-infrared, mid-infrared, and X-ray) data to perform the classifications.

The two most common ways of classifying radio AGN are by morphology and by accretion mode. The morphological method or Fanaroff-Riley classification \citet{Fanaroff-Riley1974} splits AGN based on the ratio of the distance between their regions of greatest surface brightness and the total length of the galaxy. Galaxies with their brightest regions near their centres are called FRIs while those with bright regions at the edges of extended lobes are labelled FRIIs. This is a relatively straightforward classification that can be applied when only radio images, with sufficient resolution, are present (not all AGN will fall neatly into one of these categories, making them difficult to classify). The second method, the accretion mode classification distinguishes AGN based on their radiative efficiency. Radiatively efficient AGN are those with a radiative accretion mode, meaning that they have an accretion disk and follow the orientation-based AGN unification model (\citealt{Urry1995}). They also have luminosities of $L \gtrsim 0.01L_{\textrm{Edd}}$ where $L_{\textrm{Edd}}$ is the Eddington luminosity and are therefore called high luminosity AGN (HLAGN). When these AGN are radio-loud and extended they are also called High Excitation Radio Galaxies (HERGs), but QSOs are also included in this HLAGN category. On the other hand, the radiatively inefficient AGN accrete in the jet mode, they have no accretion disk and instead their accretion flow is advection dominated. These have luminosities $L \lesssim 0.01L_{\textrm{Edd}}$ and are called Moderate luminosity AGN (MLAGN). The radio-loud variants are called Low Excitation Radio Galaxies (LERGs) and radio quiet AGN LINERS are the other members of this group. See Section 2.1 of \citet{Heckman2014AGN} for a full description of the classification. Unlike the FRI/FRII classification, this method requires a wealth of multi-wavelength data which will not always be present.

In order to conduct this multi-tracer analysis, distinct populations with different biases must be identified. The biases of SFGs have been found to be much lower than those of AGN (\citealt{hale2017}) and this is expected because AGN are hosted by more massive galaxies than the blue SFG population. We choose not to distinguish between low and high luminosity SFGs or SFG and star burst galaxies as these distinctions are difficult to make with only radio continuum data. The physically motivated choice of classification for the AGN is the accretion mode classification because MLAGN and HLAGN are expected to reside in different environments and evolve differently over time due to their different ionization properties. \citet{hale2017} found that MLAGN are significantly more biased than HLAGN and therefore inhabit more massive halos, this might be because the hot gas in the most massive halos is not easily accreted onto AGN, leading to more massive halos hosting less efficient AGN. This result supports the findings of \citealt{Hardcastle2004,Tasse2008, Janssen2012, RamosAlmeida2013, Gendre2013}. Unfortunately, the wealth of multi-wavelength data required for classifications of MLAGN/HLAGN to be made will not be present for large volumes of sky in the early stages of the SKA. On the other hand, it will be possible to identify the FRI and FRII galaxies from radio images,as long as the angular resolution is better than 1 arcsec (\citealt{Wilman2008SKADS, Ferramacho2014}). Therefore, since radio-loud MLAGN are normally associated with FRI galaxies while HLAGN are associated with FRII, for the purposes of this forecast analysis, we assume that these populations are the same although this is not strictly true. This assumption allows us to use the biases obtained from $S^3$ and \citet{hale2017} (see section \ref{bias}) and the number distributions found for FRI/FRII from the $S^3$ and T-RECS simulations (see section \ref{redshift distributions}). Therefore, in this analysis, with the aim of predicting constraints that are realistic, we use only the populations: SFGs, FRI/MLAGN and FRII/HLAGN. It is important to keep in mind the caveat that FRIs and FRIIs are similar to, but not the same as HLAGN and MLAGN respectively. For instance, the HLAGN/MLAGN distinction of the VLA-COSMOS 3 GHz survey (\citealt{Smolcic2017VLACOSMOSb}) includes the radio quiet quasars which are included as a distinct population in \citet{Wilman2008SKADS}.

\section{Fisher Analysis for SKA Forecasts} \label{fisher}

\subsection{Halo Bias} \label{bias}
Bias values for the halos of each galaxy population and for each redshift bin were required to perform the multi-tracer analysis (see equations \ref{eq:Cls} and \ref{eq:window}). In order to find these bias distributions, a mean halo mass, $M_{\rm cent}$ was adopted for each population, based on the observed/simulated estimations and it was assumed that the galaxy halos spanned a Gaussian distribution of masses, $f$, with a mean $M_{\rm cent}$ and a standard deviation of 0.2$M_{\rm cent}$. The linear halo bias distribution over a Gaussian distribution of masses, $b^i(z)$ was then calculated using the equation below:
\begin{ceqn} 
\begin{align}
b^i(z) = \int f^i(M,M^i_{\rm cent})b_L^i(M,z){\rm d}M,\label{eq:gaussian_bias}
\end{align}
\end{ceqn}
where i represents the galaxy population and $b_L^i(M,z)$ is the linear bias as a function of halo mass and redshift. \citet{Ferramacho2014} found that this simplistic approach was sufficient to describe the mass distributions and that changing this form did not notably change the constraints obtained. In order to calculate the Gaussian linear bias function above, equation \ref{eq:linearbias} was used along with the estimated average halo masses \citep[obtained from $S^3$ and][]{hale2017} and the mass variance $\sigma$ found using the Halo Mass Function (the number density of halos as a function of mass) calculator from \citet{Murray2013HMFcalc}. The linear bias is given by:
\begin{ceqn} 
\begin{align}
b_L(M,z)= 1+ \frac{q\nu -1}{\delta_c(0)} + \frac{1}{\delta_c(0)}\frac{2p}{1+(q\nu)^p},\label{eq:linearbias}
\end{align}
\end{ceqn} 
where $\nu = \delta^2_c(0)/\sigma^2(M,z)$, $\delta_c(0)$ is the critical over-density for spherical collapse at redshift $z=0$ and $\sigma(M,z)$ is the mass variance, the root mean square fluctuation in spheres which, on average, contain mass $M$ at the initial time. The accepted values of $p$ and $q$ are 0.3 and 0.75 respectively (\citealt{Sheth1999pq}). The angular power spectra code utilized in this analysis (\citealt{Fonseca2015ZeFon}) required a single bias for each redshift bin and therefore we extracted the average bias in each bin from the models found and used these values.

\begin{figure}
    \centering
    \caption{Bias redshift evolution for the different galaxy populations assuming a Gaussian distribution of masses for each population, using the central masses determined by \citet{hale2017}.}
    \includegraphics[width=0.8\columnwidth]{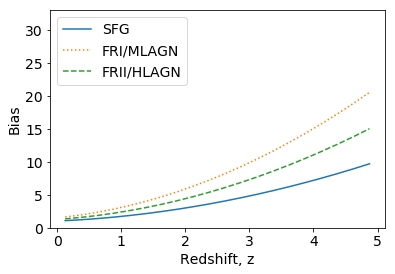}
    \label{Halebias}
\end{figure}

\begin{figure}
    \centering
    \caption{Same as Figure \ref{Halebias} using the $S^3$ simulation by \citet{Wilman2008SKADS}.}
    \includegraphics[width=0.8\columnwidth]{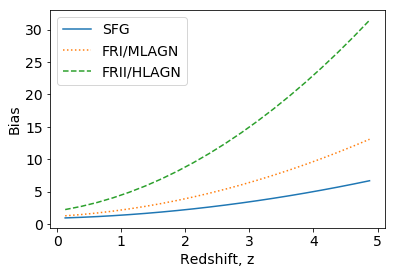}
    \label{SKADSbias}
\end{figure}

Two sets of mean halo masses were used to produce bias functions. One set of mass values were from the radio observations of \citet{hale2017} and the other was from $S^3$. The bias functions obtained for both the \citet{hale2017} and the $S^3$ masses can be seen in Figures \ref{Halebias} and \ref{SKADSbias} respectively. The bias values found by \citet{hale2017} for the MLAGN (FRI) were higher than those from $S^3$, values for the SFGs were slightly lower and values for the HLAGN (FRII) were significantly lower than those predicted by $S^3$. 

\citet{hale2017} identified one population of star forming galaxies which consisted of both normal star forming galaxies (SFGs) and star burst galaxies (SBs), while the $S^3$ simulation separated these populations. In this analysis we do not wish to distinguish between star forming and star burst galaxies as we do not expect these to be distinguished with SKA data and therefore, we choose to combine these galaxy populations of the $S^3$ simulation and used these combined biases (and redshift distributions, which will be discussed in the following section). We find the effective bias for the combined SFG and SB using a weighted mean based on the numbers of each population in each redshift bin:
\begin{ceqn} 
\begin{align}
b^{\rm eff}(z) = \frac{\sum_i n^i(z)b^i(z)}{\sum_i n^i(z)}. \label{eq:combined_bias}
\end{align}
\end{ceqn}
Following this we find the single halo mass that best fits this bias distribution and use this halo mass to conduct our analysis. In the case in which we use the $S^3$ biases and the T-RECS number distributions, we do not have differentiated SB and SFG distributions from T-RECS and therefore, we assume that the ratio of SFG to SB remains the same as in $S^3$, and we use the same bias function. The best fit halo mass was found to be $1.1 \times 10^{12} M_\odot h^{-1}$ and the corresponding bias function is plotted in Figure \ref{SKADSbias}. We notice that the combined bias of the SFG and SB of the $S^3$ simulation was similar in magnitude to the corresponding bias obtained by \citet{hale2017}.  

When exploring the constraints obtained with different populations, we also consider the case in which we are unable to differentiate the MLAGN/FRI from the HLAGN/FRII and therefore we combine the two populations, and calculate the combined bias with the method shown above. 

The bias functions presented above are extrapolated to \(z = 5\) while the bias measurements obtained from observations such as \citet{hale2017} do not extend upwards of \(z \sim 2\), \citet{Wilman2008SKADS} suggest a cut on the bias function at \(z = 1.5\) for AGN and \(z = 3\) for SFGs with a flat bias after this in order to prevent the bias from becoming unrealistically large, but there is little observational evidence for such an abrupt cut. Indeed, recent work on measuring the bias of Lyman-break galaxies at high redshifts suggest that the bias could be as high as $b\sim 8-11$ at $z\sim 6$ \citep{Hatfield2018}. This is therefore in the range of the assumed bias for our FRI/MLAGN and SFG samples, which are the dominant radio populations at these redshifts for the survey depth considered. Furthermore, one might expect that the host galaxies of the FRI/MLAGN population are more massive than the Lyman-break galaxies at similar redshifts, thus a higher bias may be expected.  However, due to the uncertainty present at redshifts above \(z \sim 2\) analyses performed with redshifts below this are more trustworthy. In this analysis we provide results for the case in which redshift is restricted to \(z<2\) and compare it to the case in which \(z<5\).

\subsection{Redshift Distributions} \label{redshift distributions}

\begin{figure}
\centering
\caption{Redshift distributions using a redshift bin size of $\Delta z = 0.25$ for the SFG, FRI and FRII populations using the $S^3$ simulation.} \label{SKADSNz}
\includegraphics[width=0.8\columnwidth]{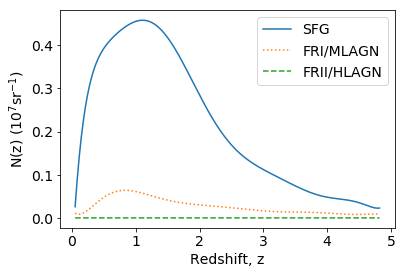}
\end{figure}

\begin{figure}
\centering
\caption{Same as Figure \ref{SKADSNz} using the T-RECS simulation.} \label{TRECSNz}
\includegraphics[width=0.8\columnwidth]{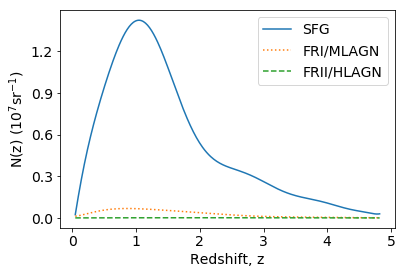}
\end{figure}

\begin{figure}
\centering
\caption{Comparison of the redshift distributions for the FRI/MLAGN (left) and FRII/HLAGN (right) populations using the $S^3$ and T-RECS simulations. A redshift bin size of $\Delta z = 0.25$ is used.} \label{FRIFRII}
\includegraphics[width=1\columnwidth]{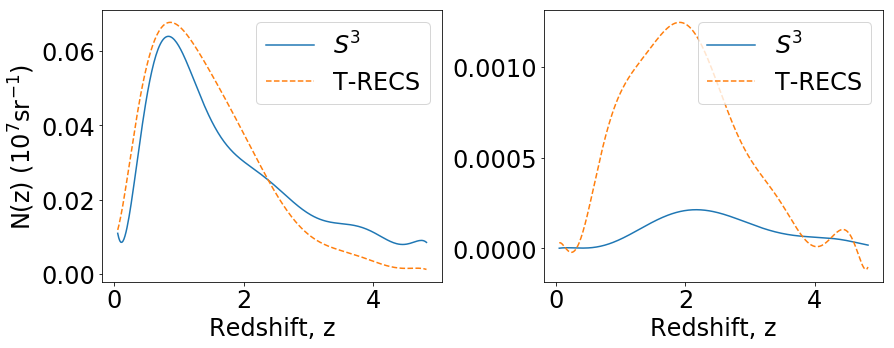}
\end{figure}

Another requirement for this analysis was redshift distributions of the galaxy populations over the redshift range of interest. Due to a lack of observed radio data at the flux limits expected for SKA (the VLA COSMOS data from \citealt{Smolcic2017VLACOSMOSa} is limited to 22 $\mu$Jy at 1.4 GHz), the redshift distributions obtained from SKA simulations, namely, the $S^3$ and T-RECS simulations were used. The MLAGN/FRI, HLAGN/FRII and SFG galaxy populations were extracted, restricted to a flux limit of 5 $\mu$Jy (a conservative flux limit for SKA Phase 1), a redshift $z <5$ and then rescaled to 1 steradian of sky. The SFG population for the $S^3$ simulation was obtained by combining the SFG and SB galaxy populations (T-RECS has one population including both of these galaxy types). The numbers of galaxies with different redshifts were then plotted in histograms using bins of size \(\Delta z = 0.25\) and functional forms were fitted. The resulting redshift distributions can be seen in Figures \ref{SKADSNz} and \ref{TRECSNz} and comparisons of the FRI and FRII distributions are shown in Figure \ref{FRIFRII}. Clearly, the T-RECS simulation predicts significantly higher numbers of SFGs than $S^3$ and these numbers seem to be more reasonable, as observations (\citealt{Smolcic2017VLACOSMOSa}) have suggested that $S^3$ underestimates SFGs (see also \citealt{Bonaldi2016}). A larger number of FRII/HLAGN was predicted by T-RECS while similar numbers of FRI/MLAGN were predicted by both simulations.

Unfortunately, since redshifts cannot be obtained from radio continuum surveys as the synchrotron radiation is featureless, it might be difficult to obtain detailed redshift distributions for large numbers of radio galaxies over large areas of sky. The emergence of large optical/near-IR surveys that will have accurate photo-$z$'s and overlapping sky coverage will mitigate this problem. On the other hand, some smaller, deep fields such as those used by the MIGHTEE survey (\citealt{Jarvis2016}) will soon have all the radio continuum, redshift and multi-wavelength data necessary for such an analysis, albeit on much smaller scales. These surveys can be used to inform the key observations necessary for the large scale $f_{\rm NL}$ studies.

The accuracy of the photometric redshifts of the survey used for cross-matching must also be sufficiently high to produce these redshift distributions, although as only distributions are necessary, precise point estimates are less important. A number of methods have been proposed for estimating redshift distributions and significant improvements on accuracy and error estimation have been made over the last few years (\citealt{Almosallam2016B, Gomes2018, Soo2018, Sanchez2019}). In addition, apart from standard template fitting, machine learning or hybrid methods, there are now techniques for determining the redshift distributions directly (without the point estimates). Two such methods are clustering redshifts (\citealt{Newman2008, Matthews2010, Menard2013, Schmidt2013}) and using magnitude space re-weighting technique (\citealt{Lima2008Nz}) both of which require a spectroscopic reference set. Most of these methods still require cross-matching the radio galaxies to a optical/near-IR survey, but the clustering redshift method does not require this as it involves the spatial cross-correlation between the galaxies with unknown redshifts and a distribution with known redshifts. This method has been shown to produce reasonable redshift distributions for radio galaxy surveys (\citealt{Menard2013}). The accuracy of the $N(z)$ distribution will become increasingly important as smaller binning schemes are chosen for analyses such as these. The redshift bin sizes of $\Delta z = 0.25$ and $\Delta z = 1$ were used in this analysis. 

An alternative option is to use the spectroscopic HI survey of the SKA but this is limited to lower redshifts $\sim 2$ and would only detect a limited sample of the star forming galaxies. Thus, it could provide a more reliable redshift distribution for the SFGs but would contribute little to the AGN number distributions.

\subsection{Fisher Analysis} \label{fisher analysis}
Fisher matrix analyses (\citealt{Tegmark1997Fisher}) allow the determination of the minimum errors on parameters that are obtainable using a particular method and experiment. This is possible because the precision to which a parameter can be determined depends on the dependence of the model on the parameter (the greater the model dependence, the greater the constraining ability) and on the covariance of the observed data (the greater the independence of the observed data, the more information can be extracted). In order to obtain the Fisher matrix for multiple tracers, in different redshift bins, we use the following expression: 
\begin{ceqn} 
\begin{align}
F_{\alpha \beta} = \sum^{l_{\rm max}}_{l=l_{\rm min}} \frac{2l+1}{2} f_{sky} Tr\Big(\frac{\partial \mathbf{C}(l)}{\partial \theta_{\alpha}}) (\mathbf{\Gamma}_l)^{-1} \frac{\partial \mathbf{C}(l)}{\partial \theta_{\beta}}) (\mathbf{\Gamma}_l)^{-1}\Big), \label{eq:fisher_matrix}
\end{align}
\end{ceqn}
where $\mathbf{C}(l)= [C^{ij}_{l}$] (defined in equation \ref{eq:Cls}) and $\theta_{\alpha}$ and $\theta_{\beta}$ are any two model parameters. $f_{sky}$ is the fraction of the sky expected to be covered by the radio survey, for this analysis we use a survey size of $2\pi$ steradian as we expect the SKA to cover at least half of the sky and therefore we use a value of $f_{sky} = 0.5$. $\mathbf{\Gamma}$ is the covariance matrix and if it is assumed that the cosmic variance and shot noise are the dominant contributors to the noise (the experimental systematics are sub-dominant) then it is given by:
\begin{ceqn} 
\begin{align}
\Gamma_l^{ij} = C_l^{ij} + \delta^{ij}N^{ii},\label{eq:gamma}
\end{align}
\end{ceqn}
where $N^{ii}$ is the noise power spectrum for a given redshift bin and galaxy population $i$ and is given by $1/n^i$ where $n^i$ is the number of sources per steradian. These values of $n^i$ were obtained using the scaled redshift distributions discussed above. 

The bias and number distributions were then used to calculate the auto and cross correlation power spectra using equations \ref{eq:Cls} and \ref{eq:window}. For this computation a modification of CAMB sources (\citealt{Challinor2011CAMBsources}) presented by \citet{Fonseca2015ZeFon} was used. This code allows the user to input the redshift distributions and biases of the various tracers. In the most ideal case we split the redshift distribution into top-hat redshift bins of width 0.25 (a conservative size as we expect the photo-$z$ estimates/spectroscopic redshifts to be much more precise than this) leading to a total of 20 windows for each of the three galaxy populations. The fiducial parameters were then inputted and power spectra were obtained with $l_{\rm min} = 2$ and $l_{\rm max} = 200$ (as in \citealt{Ferramacho2014}). The cosmological parameters that are most degenerate with the non-Gaussianity signal are dark matter density $\Omega_{\textrm{CDM}}$, the Hubble parameter $H_0$ and the amplitude of the primordial fluctuations $A_s$ and therefore these are included in the Fisher analysis. We use the parameters obtained by Planck 2015 (\citealt{Planck2016PNG}) with CMB temperature and polarization power spectra as well as lensing and external (Baryon Acoustic Oscillations, supernova Joint Light-curve Analysis and $H_0$) measurements. The other cosmological parameters were used in producing the angular power spectra. The parameters used for this analysis (the cosmological parameters and the estimated central masses of the galaxy populations) and their fiducial values are given in Table \ref{tab:fiducial_variables}. 

\begin{table}
\centering
\begin{tabular}{p{2cm}p{2cm}p{2cm}}
  \toprule
Parameter &  \multicolumn{2}{c}{Fiducial Value} \\ 
  \hline

 $\Omega_{\textrm{CDM}}$  & \multicolumn{2}{c}{0.26}  \\
 $H_0$      & \multicolumn{2}{c}{67.74}  \\
 $ln(10^{10}A_s)$  & \multicolumn{2}{c}{3.064}\\
 $f_{\rm NL}$  & \multicolumn{2}{c}{0} \\
\hline
    &\raggedright Hale Halo Masses & $S^3$ Halo Masses \\
\hline

 \addlinespace[0.1cm]
 $M_{\rm cent}^{\textrm{SFG}}$  & $4 \times 10^{12}$  & $1.1 \times 10^{12}$ \\
 \addlinespace[0.2cm]
 $M_{\rm cent}^{\textrm{MLAGN}}$ & $3.5 \times 10^{13}$  & $1 \times 10^{13}$ \\
 \addlinespace[0.2cm] 
 $M_{\rm cent}^{\textrm{HLAGN}}$ & $1.5 \times 10^{13}$  & $1 \times 10^{14}$  \\
 \addlinespace[0.1cm]
 $M_{\rm cent}^{\textrm{AGN}}$ & $3.46 \times 10^{13}$  & $1.01 \times 10^{14}$  \\
 \addlinespace[0.1cm]
\bottomrule
\end{tabular}
\caption{Fiducial values chosen for the parameters in this model. The masses are in units of $M_\odot h^{-1}$,  where $h = H_0/100$.}
 \label{tab:fiducial_variables}
\end{table}

The next step was to obtain the derivatives of these power spectra with respect to each parameter. Some of the derivatives were computed analytically while others were computed numerically using a 5-point stencil method. The covariance matrix, $\Gamma$, was then computed using the $C_l$ results with the fiducial cosmology according to equation \ref{eq:gamma}. The fisher matrix was then computed using equation \ref{eq:fisher_matrix}. The constraint on the $f_{\rm NL}$ parameter was obtained by inverting the Fisher matrix and extracting the $f_{\rm NL}$ $\times$ $f_{\rm NL}$ component.

\section{SKA Forecasts - Results and Analysis} \label{results}

In order to understand the relationships between the constraints that can be obtained with differently sized redshift bins (corresponding to photometric redshift quality) and different redshift ranges (which will depend on the performance of the SKA as well as the availability of multi-wavelength data with photometric redshifts to cross match to the radio data) we considered a number of cases:
\begin{itemize}
    \item Redshift bins of width 0.25 and a range of $0<z<5$
    \item Redshift bins of width 1 and a range of $0<z<5$
    \item Redshift bins of width 0.25 and a range of $0<z<2$
\end{itemize}
The first of these cases represents the most optimistic view, implying that there will be complete and accurate photometric redshift coverage to match to the radio galaxies up to a redshift of $z \sim 5$. This is unlikely to be the case due to obstacles such as dust obscuration at high redshifts which could introduce artificial selection effects (with brighter objects more likely to be detected) and impact our measurements of $f_{\rm NL}$. In addition, there are limited spectroscopic redshifts at high redshifts, making the training of machine learning photometric redshift estimation methods as well as the redshift distribution estimation methods more difficult and less accurate. For these reasons, we also evaluated the case in which we are able to get redshifts up to 5 but we have much lower photometric redshift accuracy and therefore we use much larger bins of width 1. Finally we also consider the much more realistic case of obtaining galaxies with photometric redshifts only up to a redshift of 2. Below this redshift it is more likely that multi-wavelength data, and correspondingly photometric redshifts will be available for cross matching to most of the radio sources. In addition, in this redshift range it is also possible to obtain some redshifts spectroscopically from HI line measurements. Photometric redshift accuracy is also expected to be higher in this redshift range, making the bin width of 0.25 a conservative choice. In addition to obtaining better redshifts below \(z \sim 2\), the bias functions measured/simulated are also more accurate below these redshifts. As a result, constraints obtained for this case of restricted redshift are much more realistic than other cases.

For each of these cases we evaluated the constraints on $f_{\rm NL}$ with the different combinations of either the $S^3$ or \citet{hale2017} biases and the $S^3$ or T-RECS $N(z)$ distributions in order to obtain a range of constraints depending on the range of possibilities for the radio source populations. We note that combining the $N(z)$ and the biases, from different studies and techniques, in this way could be viewed as internally inconsistent. However, we believe that this gives an indication of how the different assumed distributions effect the results.
We also compared the result of having 3 independent distributions: SFG+SB, MLAGN/FRI and HLAGN/FRII to the case in which MLAGN/FRI and HLAGN/FRII are not distinguished and we have 2 populations: SFG+SB and AGN. The morphology of radio galaxies is not a discrete distribution of FRI and FRII objects, but a continuous distribution of morphologies, with many objects that are difficult to classify as either an FRI or a FRII. This distinction also becomes increasingly difficult as redshift increases and the images obtained resolve less structure. Therefore, for the most conservative case we assume that we are not able to make this distinction. We also observed that the number density of the HLAGN in both redshift distributions was relatively low in comparison to the other populations and therefore we decided to compare the constraints obtained when we use just SFG+SB and MLAGN to SFG+SB and AGN (where the HLAGN is included in the AGN).  

\begin{table*}
\centering
\begin{tabular}{p{1.5cm}p{1.5cm}p{2cm}}
  \toprule
 & $S^3$ $N(z)$ & T-RECS $N(z)$ \\ 
  \hline
 $S^3$ biases      & 1.90 & 1.88 \\
\addlinespace[0.1cm]
 Hale biases      & 1.54 & 1.56 \\
\bottomrule
\end{tabular}
\caption{Forecasts on $f_{\rm NL}$ $1-\sigma$ errors obtained using a multi-tracer Fisher analysis with redshift bins of size $\Delta z= 0.25$ and a range of $0<z<5$.}
 \label{tab:results_small_bins_large_range}
\end{table*}

\begin{table*}
\centering
\begin{tabular}{p{2cm}p{2cm}p{4cm}p{3cm}p{3cm}}
  \toprule
  & & SFG+SB MLAGN HLAGN  & SFG+SB MLAGN & SFG+SB AGN  \\ 
  \hline
$S^3$ biases & $S^3$ $N(z)$  & 2.23 & 2.27 & 2.26 \\
  \addlinespace[0.1cm]
  & T-RECS $N(z)$      & 2.35 & 2.47 & 2.46 \\
  \addlinespace[0.1cm]
   Hale biases & $S^3$ $N(z)$  & 1.83 & 1.84 & 1.84 \\
   \addlinespace[0.1cm]
   & T-RECS $N(z)$      & 1.98 & 1.99 & 1.98 \\
\bottomrule
\end{tabular}
\caption{Forecasts on $f_{\rm NL}$ $1-\sigma$ errors obtained using a multi-tracer Fisher analysis with redshift bins of size $\Delta z = 1$ and a range of $0<z<5$.}
 \label{tab:results_large_bins_large_range}
\end{table*}

\begin{table*}
\centering
\begin{tabular}{p{2cm} p{2cm} p{4cm} p{3cm}}
  \toprule
   &  & SFG+SB MLAGN HLAGN & SFG+SB AGN \\ 
  \hline
$S^3$ biases & $S^3$ $N(z)$  & 6.53 & 6.58 \\
  \addlinespace[0.1cm]
  & T-RECS $N(z)$ & 5.39 & 5.82 \\
  \addlinespace[0.1cm]
   Hale biases & $S^3$ $N(z)$  & 4.61 & 4.63 \\
   \addlinespace[0.1cm]
   & T-RECS $N(z)$      & 4.07 & 4.08 \\
\bottomrule
\end{tabular}
\caption{Forecasts on $f_{\rm NL}$ $1-\sigma$ errors obtained using a multi-tracer Fisher analysis with redshift bins of size $\Delta z = 0.25$ and a range of $0<z<2$.}
 \label{tab:results_small_bins_small_range}
\end{table*}

The constraints measured for each of the cases presented above are shown in Tables \ref{tab:results_small_bins_large_range}, \ref{tab:results_large_bins_large_range} and \ref{tab:results_small_bins_small_range}. We found that for all cases (of redshift bin size and range as well as combinations of galaxy types) the biases from \citet{hale2017} lead to significantly tighter constraints than those from $S^3$, while the different redshift distributions of the two simulations lead to less notable differences. This means that if the biases calculated by \citet{hale2017} are nearer to the truth, tighter constraints than previously expected (\citealt{Ferramacho2014}) will be possible with this method. The \citet{hale2017} biases may have caused this effect because the separation in the biases of the FRI and SFGs is much greater than for $S^3$ (compare Figures \ref{Halebias} and \ref{SKADSbias}) and the numbers of SFG and FRI are much greater than the number of FRII present. Therefore, although the difference between the bias functions of the populations is greater for $S^3$, with the FRII galaxies having very high biases, the very limited number of FRIIs and larger numbers of FRIs and SFGs lead to the \citet{hale2017} biases producing tighter constraints. We note that ideally, the \(N(z)\) distribution parameters would be marginalized over in our analysis.  However, the differences in the constraints caused by the \(N(z)\) distributions is subdominant to the differences due to the biases using the $S^3$ simulation and \citet{hale2017}. Therefore, the fact that the number distributions have substantially different distributions (see Figures \ref{SKADSNz} and \ref{TRECSNz}) implies that small variations in the number distributions will not result in notable differences in constraints.

 We observed that the effect of the $S^3$ versus the T-RECS redshift distributions varied across the redshift bin size and range. For the case with $z<2$ (Table \ref{tab:results_small_bins_small_range}), the T-RECS number distributions produced significantly tighter constraints while for the case of larger redshift bins but a range of $0<z<5$ (Table \ref{tab:results_large_bins_large_range}) the $S^3$ number distributions resulted in tighter constraints. This pattern is likely because the numbers of MLAGN in the T-RECS simulation tapers off to much lower numbers beyond $z = 2$. This causes the MLAGN in the $S^3$ simulation to contribute more to the $0<z<5$ measurement than those in the T-RECS simulation, while the MLAGN in the T-RECS simulation contributes more to the $z<2$ measurement than those in the $S^3$ simulation. On the other hand, in the most ideal case with small redshift bins and a large redshift range the differences induced by the different redshift distributions are insignificant. 

Excluding the HLAGN (see Table \ref{tab:results_large_bins_large_range}) and joining the MLAGN and HLAGN (Tables \ref{tab:results_large_bins_large_range} and \ref{tab:results_small_bins_small_range}) both reduced the constraints (increased the $1 - \sigma$ errors) by small but notable amounts with the combined MLAGN and HLAGN performing marginally better than removing the HLAGN entirely. This is expected as the HLAGN population is very small and therefore does not contribute much to the constraints when added to the MLAGN. The addition of the HLAGN does increase the numbers to a small extent, which decreases the shot noise, while also contributing a greater ratio of higher redshift galaxies, these effects lead to the marginal improvement observed. On the other hand, removing the HLAGN as an independent tracer does decrease the constraints notably as there are now only two tracers with different biases to be compared. This reduction in constraining power was found to be much larger for the combination of the $S^3$ biases and the T-RECS $N(z)$ distributions than for the other cases and this is due to the very large bias values assigned to the HLAGN by the $S^3$ simulation coupled with the larger numbers of HLAGN predicted by the T-RECS simulation. Another important observation is that the increases in errors imposed by the removal of the HLAGN as an independent tracer are much less significant than the variation between the bias and number distribution configurations. This is clearly observed in Tables \ref{tab:results_large_bins_large_range} and \ref{tab:results_small_bins_small_range} and proves that the SFG and MLAGN populations are contributing the majority of the constraining power. This implies that if image resolution or multi-wavelength data is not sufficient for distinguishing between MLAGN/FRI and HLAGN/FRII then this is not a major problem as it will not significantly decrease the possible constraints.                                          

When comparing the cases with different redshift bin widths we find that increasing the bin size increased the $1 \sigma$ errors on $f_{\rm NL}$ on all the bias and $N(z)$ configurations by a factor less than 1.5 (compare Table \ref{tab:results_small_bins_large_range} to Table \ref{tab:results_large_bins_large_range}) while reducing the redshift range to $z<2$ had a much more significant impact with a factor nearer to 3 (see Table \ref{tab:results_small_bins_large_range} and Table \ref{tab:results_small_bins_small_range}). This implies that while having accurate photometric redshift estimates that allow smaller bins is important it is not as vital to adding constraining power as having deeper surveys with galaxies and cross-matched photometric redshifts going to higher redshifts. The reduced redshift led to more significant reductions in constraining power because the volume of the survey is decreased, thus reducing the number of large scale measurements contributing to the angular power spectra. Despite this, as previously discussed, we do not expect to have these data at $z\gtrsim 2$ in the near future, and the bias functions have not been measured at higher redshifts, therefore, we take the results in Table \ref{tab:results_small_bins_small_range} as the expected limits of constraints with the SKA Phase 1 continuum surveys. Note here that our choice of $f_{sky} = 0.5$ was conservative for the SKA and it is possible that greater coverage will be obtained, as the errors scale with $\sqrt{ f_{sky}}$ this will lead to tighter constraints. Also note that these fisher matrix constraints are the minimum possible errors and that their calculation assumes that any systematic errors inherent in the experiment have a much smaller impact than the cosmic variance or shot noise contributions.   

It should be noted that this was a conservative analysis, using only the three or two galaxy populations that are most likely to be correctly distinguished. If the multi-wavelength data available becomes sufficient to distinguish star burst galaxies from the other star forming galaxies and to similarly differentiate the radio quiet quasars then additional tracers can be utilized and the results will be further improved, as evaluated in \citet{Ferramacho2014}.

\section{Conclusions} \label{conclusions}
In this paper we have presented an analysis of the constraints on primordial non-Gaussianity that could be obtained using the SKA radio continuum surveys. This paper serves as an update to the analysis undertaken by \citet{Ferramacho2014}. We investigated the possibility of using multiple radio galaxy populations, with different halo mass and bias properties to perform a multi-tracer analysis, thus reducing the effect of cosmic variance and increasing the constraining power that can be obtained. The galaxy populations we considered were star forming galaxies, moderate luminosity AGN (or FRIs) and high luminosity AGN (or FRIIs). We consider the MLAGN and HLAGN to be similar enough to the FRI and FRII populations that we can use the classifications interchangeably based on the simulation/observations utilized. In contrast to \citet{Ferramacho2014} we do not distinguish star forming galaxies and star burst galaxies as this may not be observationally feasible at the required level of precision.

We perform a Fisher matrix analysis following the formalism of \citet{Ferramacho2014} which utilizes the auto- and cross-correlation angular power spectra of the multi-tracers. These power spectra require the redshift and bias distributions of the radio galaxy populations. We use the redshift distributions of radio galaxies populations given by two SKA simulations: the $S^3$ simulation and the more recent T-RECS simulation and we re-scaled them to the appropriate area and flux limit. For the bias distributions we used the average halo masses obtained from \citet{hale2017} who use VLA COSMOS-field observations as well as those estimated by $S^3$. We then assumed a Gaussian distribution around this mass and calculated the bias using an analytical expression for the linear bias. We considered all of the combinations of the bias and redshift distributions as well as different redshift ranges (dependent on the depth of the continuum survey as well as the availability of visible/near-IR data with reliable photometric redshifts that can be cross-matched to the radio sources) and redshift bin sizes (dependent on photometric redshift accuracy). In addition to treating the SFG, MLAGN/FRI and HLAGN/FRII as independent populations we also considered the combined MLAGN and HLAGN population to determine the effect on the constraints if the MLAGN and HLAGN cannot be differentiated. We take the case of small redshift bins (width = 0.25) and a small redshift range ($0<z<2$) as the most realistic case for SKA observations and this provides a range in the 1-$\sigma$ errors of 4.07 to 6.53 if the AGN populations are differentiated and 4.08 to 6.58 if they are not, the similarity between these is due to the fact that the high-luminosity AGN are much rarer than the dominant AGN population at lower luminosities. The results with the observed biases (\citealt{hale2017}) and the more recent simulations (T-RECS) provide the lower bounds of these ranges: 4.07 and 4.08 with 3 populations and 2 populations respectively. These constraints will surpass the existing tightest constraints \citep[$\sim 5$ obtained with Planck 2015 and 2018 data; ][]{Planck2016PNG,Planck2018PNG} but will also provide an independent precise measure of $f_{\rm NL}$, obtained from the large-scale structure instead of the CMB bispectrum. These results also indicate that if redshift information is available for the galaxy populations to higher redshifts, constraints will improve significantly, and with a redshift range of $0<z<5$ the 1$\sigma$ error on $f_{\rm NL}$ will fall between 1.5 and 2, representing unprecedented precision on the measurement of $f_{\rm NL}$ and approaching the target of 1. On the other hand, some limitations of this analysis is that the masses of the galaxy populations might have non-Gaussian distributions or different standard deviations than the ones assumed in this analysis, in addition, the bias evolution used in this analysis is very uncertain at redshifts above \(z \sim 2\). Moreover, the redshift distributions used are based on simulations, which are constrained by observations, but they still might not be very accurate, particularly at the higher redshifts where the lever-arm for the constraints on $f_{\rm NL}$ are strongest.

It is clear that in order to use wide-area radio continuum surveys for constraining the influence of non-Gaussianity on the the large-scale structure, then robust measurements of the distribution of biases for the different populations, along with their redshift distributions are needed. To some extent, this is underway, as we have used the observed constraints from the VLA-COSMOS survey in order to estimate the bias of the variety of radio source populations. However, VLA-COSMOS covers a relatively small area, and does not reach the depth of the planned ``all-sky" SKA continuum surveys. Therefore, the imminent MeerKAT International Giga-Hertz Tiered Extragalactic Exploration \citep[MIGHTEE; ][]{Jarvis2016} survey will certainly provide much better information on the bias (through a clustering analysis covering many square degrees, as opposed to a single $\sim 2$\,deg field, and to a depth similar to that of the all-sky SKA survey. A key element of MIGHTEE is it also covers areas of the extragalactic sky with excellent multi-wavelength data, thus more accurate redshift distributions will also be available based on the photometric redshifts in these fields (similar to VLA-COSMOS now). 

The combination of these multi-wavelength data, coupled with deep radio data may also allow the characterisation of the sources into the populations discussed here using morphological information. Such an effort could potentially supply the necessary training sample for a variety of machine learning algorithms currently being tested in radio data \citep[e.g][]{Lukic2018,Glaser2019}, and then applied to wider field data where the data extent and quality is not comparable to in these deep fields.

Thus, although constraining $f_{\rm NL}$ requires large cosmological volumes, the information that can be gleaned from the deep narrow surveys will be crucial in planning the final strategy.

Furthermore, joint analyses with other multi-tracers probes (other than radio continuum) such as HI intensity mapping or low redshift galaxy surveys would likely lead to further improvements on $f_{\rm NL}$ constraints.

\section*{Acknowledgements}
ZG is supported by a Rhodes Scholarship granted by the Rhodes Trust.  MJJ acknowledges support from the South African Radio Astronomy Observatory through a visiting Professorship. CLH would like to
acknowledge the support given from the Science and Technology Facilities Council (STFC) through
an STFC studentship. SC is supported by the Italian Ministry of Education, University and Research (MIUR) through Rita Levi
Montalcini project `\textsc{prometheus} - Probing and Relating Observables with Multi-wavelength Experiments To
Help Enlightening the Universe's Structure', and by the
`Departments of Excellence 2018-2022' Grant awarded by MIUR (L. 232/2016). JF acknowledges support from the South African Square Kilometre Array Project and National Research Foundation.

\bibliographystyle{mnras}
\bibliography{refs_png}

\begin{thebibliography}{}
\makeatletter
\relax
\def\mn@urlcharsother{\let\do\@makeother \do\$\do\&\do\#\do\^\do\_\do\%\do\~}
\def\mn@doi{\begingroup\mn@urlcharsother \@ifnextchar [ {\mn@doi@}
  {\mn@doi@[]}}
\def\mn@doi@[#1]#2{\def\@tempa{#1}\ifx\@tempa\@empty \href
  {http://dx.doi.org/#2} {doi:#2}\else \href {http://dx.doi.org/#2} {#1}\fi
  \endgroup}
\def\mn@eprint#1#2{\mn@eprint@#1:#2::\@nil}
\def\mn@eprint@arXiv#1{\href {http://arxiv.org/abs/#1} {{\tt arXiv:#1}}}
\def\mn@eprint@dblp#1{\href {http://dblp.uni-trier.de/rec/bibtex/#1.xml}
  {dblp:#1}}
\def\mn@eprint@#1:#2:#3:#4\@nil{\def\@tempa {#1}\def\@tempb {#2}\def\@tempc
  {#3}\ifx \@tempc \@empty \let \@tempc \@tempb \let \@tempb \@tempa \fi \ifx
  \@tempb \@empty \def\@tempb {arXiv}\fi \@ifundefined
  {mn@eprint@\@tempb}{\@tempb:\@tempc}{\expandafter \expandafter \csname
  mn@eprint@\@tempb\endcsname \expandafter{\@tempc}}}

\bibitem[\protect\citeauthoryear{{Almosallam}, {Jarvis}  \&
  {Roberts}}{{Almosallam} et~al.}{2016}]{Almosallam2016B}
{Almosallam} I.~A.,  {Jarvis} M.~J.,   {Roberts} S.~J.,  2016, \mn@doi [\mnras]
  {10.1093/mnras/stw1618}, \href
  {http://adsabs.harvard.edu/abs/2016MNRAS.462..726A} {462, 726}

\bibitem[\protect\citeauthoryear{{Alonso} \& {Ferreira}}{{Alonso} \&
  {Ferreira}}{2015}]{Alonso2015redblue}
{Alonso} D.,  {Ferreira} P.~G.,  2015, \mn@doi [\prd]
  {10.1103/PhysRevD.92.063525}, \href
  {https://ui.adsabs.harvard.edu/abs/2015PhRvD..92f3525A} {92, 063525}

\bibitem[\protect\citeauthoryear{{Alvarez} et~al.,}{{Alvarez}
  et~al.}{2014}]{Alvarez2014}
{Alvarez} M.,  et~al., 2014, arXiv e-prints, \href
  {https://ui.adsabs.harvard.edu/abs/2014arXiv1412.4671A} {p. arXiv:1412.4671}

\bibitem[\protect\citeauthoryear{{Ballardini}, {Matthewson}  \&
  {Maartens}}{{Ballardini} et~al.}{2019}]{Ballardini2019}
{Ballardini} M.,  {Matthewson} W.~L.,   {Maartens} R.,  2019, arXiv e-prints,
  \href {https://ui.adsabs.harvard.edu/abs/2019arXiv190604730B} {p.
  arXiv:1906.04730}

\bibitem[\protect\citeauthoryear{{Bartolo}, {Komatsu}, {Matarrese}  \&
  {Riotto}}{{Bartolo} et~al.}{2004}]{Bartolo2004Inf}
{Bartolo} N.,  {Komatsu} E.,  {Matarrese} S.,   {Riotto} A.,  2004, \mn@doi
  [\physrep] {10.1016/j.physrep.2004.08.022}, \href
  {http://adsabs.harvard.edu/abs/2004PhR...402..103B} {402, 103}

\bibitem[\protect\citeauthoryear{{Bonaldi}, {Harrison}, {Camera}  \&
  {Brown}}{{Bonaldi} et~al.}{2016}]{Bonaldi2016}
{Bonaldi} A.,  {Harrison} I.,  {Camera} S.,   {Brown} M.~L.,  2016, \mn@doi
  [\mnras] {10.1093/mnras/stw2104}, \href
  {https://ui.adsabs.harvard.edu/abs/2016MNRAS.463.3686B} {463, 3686}

\bibitem[\protect\citeauthoryear{{Bonaldi}, {Bonato}, {Galluzzi}, {Harrison},
  {Massardi}, {Kay}, {De Zotti}  \& {Brown}}{{Bonaldi}
  et~al.}{2018}]{Bonaldi2018TRECS}
{Bonaldi} A.,  {Bonato} M.,  {Galluzzi} V.,  {Harrison} I.,  {Massardi} M.,
  {Kay} S.,  {De Zotti} G.,   {Brown} M.~L.,  2018, preprint, \href
  {http://adsabs.harvard.edu/abs/2018arXiv180505222B} {} (\mn@eprint {arXiv}
  {1805.05222})

\bibitem[\protect\citeauthoryear{Bruni, Crittenden, Koyama, Maartens, Pitrou
  \& Wands}{Bruni et~al.}{2012}]{Bruni2012}
Bruni M.,  Crittenden R.,  Koyama K.,  Maartens R.,  Pitrou C.,   Wands D.,
  2012, \mn@doi [Phys. Rev. D] {10.1103/PhysRevD.85.041301}, 85, 041301

\bibitem[\protect\citeauthoryear{{Camera}, {Santos}, {Bacon}, {Jarvis},
  {McAlpine}, {Norris}, {Raccanelli}  \& {R{\"o}ttgering}}{{Camera}
  et~al.}{2012}]{Camera2012}
{Camera} S.,  {Santos} M.~G.,  {Bacon} D.~J.,  {Jarvis} M.~J.,  {McAlpine} K.,
  {Norris} R.~P.,  {Raccanelli} A.,   {R{\"o}ttgering} H.,  2012, \mn@doi
  [\mnras] {10.1111/j.1365-2966.2012.22073.x}, \href
  {https://ui.adsabs.harvard.edu/abs/2012MNRAS.427.2079C} {427, 2079}

\bibitem[\protect\citeauthoryear{{Camera}, {Santos}, {Ferreira}  \&
  {Ferramacho}}{{Camera} et~al.}{2013}]{Camera2013IM}
{Camera} S.,  {Santos} M.~G.,  {Ferreira} P.~G.,   {Ferramacho} L.,  2013,
  \mn@doi [Physical Review Letters] {10.1103/PhysRevLett.111.171302}, \href
  {http://adsabs.harvard.edu/abs/2013PhRvL.111q1302C} {111, 171302}

\bibitem[\protect\citeauthoryear{{Camera}, {Maartens}  \& {Santos}}{{Camera}
  et~al.}{2015}]{Camera2015GR}
{Camera} S.,  {Maartens} R.,   {Santos} M.~G.,  2015, \mn@doi [\mnras]
  {10.1093/mnrasl/slv069}, \href
  {https://ui.adsabs.harvard.edu/abs/2015MNRAS.451L..80C} {451, L80}

\bibitem[\protect\citeauthoryear{{Carbone}, {Verde}  \& {Matarrese}}{{Carbone}
  et~al.}{2008}]{Carbone2008LSS}
{Carbone} C.,  {Verde} L.,   {Matarrese} S.,  2008, \mn@doi [\apjl]
  {10.1086/592020}, \href {http://adsabs.harvard.edu/abs/2008ApJ...684L...1C}
  {684, L1}

\bibitem[\protect\citeauthoryear{Carilli \& Rawlings}{Carilli \&
  Rawlings}{2004}]{Carilli2004}
Carilli C.,  Rawlings S.,  2004, \mn@doi [New Astronomy Reviews]
  {https://doi.org/10.1016/j.newar.2004.09.001}, 48, 979

\bibitem[\protect\citeauthoryear{{Castorina}, {Feng}, {Seljak}  \&
  {Villaescusa-Navarro}}{{Castorina} et~al.}{2018}]{Castorina2018zerobias}
{Castorina} E.,  {Feng} Y.,  {Seljak} U.,   {Villaescusa-Navarro} F.,  2018,
  preprint, \href {http://adsabs.harvard.edu/abs/2018arXiv180311539C} {}
  (\mn@eprint {arXiv} {1803.11539})

\bibitem[\protect\citeauthoryear{{Castorina} et~al.,}{{Castorina}
  et~al.}{2019}]{Castorina2019}
{Castorina} E.,  et~al., 2019, \mn@doi [\jcap] {10.1088/1475-7516/2019/09/010},
  \href {https://ui.adsabs.harvard.edu/abs/2019JCAP...09..010C} {2019, 010}

\bibitem[\protect\citeauthoryear{{Challinor} \& {Lewis}}{{Challinor} \&
  {Lewis}}{2011}]{Challinor2011CAMBsources}
{Challinor} A.,  {Lewis} A.,  2011, \mn@doi [\prd]
  {10.1103/PhysRevD.84.043516}, \href
  {http://adsabs.harvard.edu/abs/2011PhRvD..84d3516C} {84, 043516}

\bibitem[\protect\citeauthoryear{{Creminelli} \& {Zaldarriaga}}{{Creminelli} \&
  {Zaldarriaga}}{2004}]{Creminelli2004}
{Creminelli} P.,  {Zaldarriaga} M.,  2004, \mn@doi [Journal of Cosmology and
  Astro-Particle Physics] {10.1088/1475-7516/2004/10/006}, \href
  {https://ui.adsabs.harvard.edu/\#abs/2004JCAP...10..006C} {2004, 006}

\bibitem[\protect\citeauthoryear{{Dalal}, {Dor{\'e}}, {Huterer}  \&
  {Shirokov}}{{Dalal} et~al.}{2008}]{Dalal2008LSS}
{Dalal} N.,  {Dor{\'e}} O.,  {Huterer} D.,   {Shirokov} A.,  2008, \mn@doi
  [\prd] {10.1103/PhysRevD.77.123514}, \href
  {http://adsabs.harvard.edu/abs/2008PhRvD..77l3514D} {77, 123514}

\bibitem[\protect\citeauthoryear{De~Bernardis, Serra, Cooray  \&
  Melchiorri}{De~Bernardis et~al.}{2010}]{Bernardis2010LRG}
De~Bernardis F.,  Serra P.,  Cooray A.,   Melchiorri A.,  2010, \mn@doi [Phys.
  Rev. D] {10.1103/PhysRevD.82.083511}, 82, 083511

\bibitem[\protect\citeauthoryear{{Dewdney}, {Hall}, {Schilizzi}  \&
  {Lazio}}{{Dewdney} et~al.}{2009}]{Dewdney2009}
{Dewdney} P.~E.,  {Hall} P.~J.,  {Schilizzi} R.~T.,   {Lazio} T.~J.~L.~W.,
  2009, \mn@doi [IEEE Proceedings] {10.1109/JPROC.2009.2021005}, \href
  {http://adsabs.harvard.edu/abs/2009IEEEP..97.1482D} {97, 1482}

\bibitem[\protect\citeauthoryear{{Dor{\'e}} et~al.,}{{Dor{\'e}}
  et~al.}{2014}]{Dore2014}
{Dor{\'e}} O.,  et~al., 2014, arXiv e-prints, \href
  {https://ui.adsabs.harvard.edu/\#abs/2014arXiv1412.4872D} {p.
  arXiv:1412.4872}

\bibitem[\protect\citeauthoryear{{Fanaroff} \& {Riley}}{{Fanaroff} \&
  {Riley}}{1974}]{Fanaroff-Riley1974}
{Fanaroff} B.~L.,  {Riley} J.~M.,  1974, \mn@doi [\mnras]
  {10.1093/mnras/167.1.31P}, \href
  {http://adsabs.harvard.edu/abs/1974MNRAS.167P..31F} {167, 31P}

\bibitem[\protect\citeauthoryear{{Ferramacho}, {Santos}, {Jarvis}  \&
  {Camera}}{{Ferramacho} et~al.}{2014}]{Ferramacho2014}
{Ferramacho} L.~D.,  {Santos} M.~G.,  {Jarvis} M.~J.,   {Camera} S.,  2014,
  \mn@doi [\mnras] {10.1093/mnras/stu1015}, \href
  {http://adsabs.harvard.edu/abs/2014MNRAS.442.2511F} {442, 2511}

\bibitem[\protect\citeauthoryear{{Ferraro} \& {Smith}}{{Ferraro} \&
  {Smith}}{2015}]{Ferraro2015}
{Ferraro} S.,  {Smith} K.~M.,  2015, \mn@doi [\prd]
  {10.1103/PhysRevD.91.043506}, \href
  {https://ui.adsabs.harvard.edu/abs/2015PhRvD..91d3506F} {91, 043506}

\bibitem[\protect\citeauthoryear{{Fonseca}, {Camera}, {Santos}  \&
  {Maartens}}{{Fonseca} et~al.}{2015}]{Fonseca2015ZeFon}
{Fonseca} J.,  {Camera} S.,  {Santos} M.~G.,   {Maartens} R.,  2015, \mn@doi
  [\apjl] {10.1088/2041-8205/812/2/L22}, \href
  {http://adsabs.harvard.edu/abs/2015ApJ...812L..22F} {812, L22}

\bibitem[\protect\citeauthoryear{{Fonseca}, {Maartens}  \& {Santos}}{{Fonseca}
  et~al.}{2017}]{Fonseca2017}
{Fonseca} J.,  {Maartens} R.,   {Santos} M.~G.,  2017, \mn@doi [\mnras]
  {10.1093/mnras/stw3248}, \href
  {https://ui.adsabs.harvard.edu/\#abs/2017MNRAS.466.2780F} {466, 2780}

\bibitem[\protect\citeauthoryear{{Fonseca}, {Maartens}  \& {Santos}}{{Fonseca}
  et~al.}{2018}]{Fonseca2018}
{Fonseca} J.,  {Maartens} R.,   {Santos} M.~G.,  2018, \mn@doi [\mnras]
  {10.1093/mnras/sty1702}, \href
  {https://ui.adsabs.harvard.edu/\#abs/2018MNRAS.479.3490F} {479, 3490}

\bibitem[\protect\citeauthoryear{{Gendre}, {Best}, {Wall}  \& {Ker}}{{Gendre}
  et~al.}{2013}]{Gendre2013}
{Gendre} M.~A.,  {Best} P.~N.,  {Wall} J.~V.,   {Ker} L.~M.,  2013, \mn@doi
  [\mnras] {10.1093/mnras/stt116}, \href
  {https://ui.adsabs.harvard.edu/#abs/2013MNRAS.430.3086G} {430, 3086}

\bibitem[\protect\citeauthoryear{{Giannantonio} \& {Percival}}{{Giannantonio}
  \& {Percival}}{2014}]{Giannantonio2014}
{Giannantonio} T.,  {Percival} W.~J.,  2014, \mn@doi [\mnras]
  {10.1093/mnrasl/slu036}, \href
  {http://adsabs.harvard.edu/abs/2014MNRAS.441L..16G} {441, L16}

\bibitem[\protect\citeauthoryear{{Glaser}, {Wong}, {Schawinski}  \&
  {Zhang}}{{Glaser} et~al.}{2019}]{Glaser2019}
{Glaser} N.,  {Wong} O.~I.,  {Schawinski} K.,   {Zhang} C.,  2019, \mn@doi
  [\mnras] {10.1093/mnras/stz1534}, \href
  {https://ui.adsabs.harvard.edu/abs/2019MNRAS.487.4190G} {487, 4190}

\bibitem[\protect\citeauthoryear{{Gomes}, {Jarvis}, {Almosallam}  \&
  {Roberts}}{{Gomes} et~al.}{2018}]{Gomes2018}
{Gomes} Z.,  {Jarvis} M.~J.,  {Almosallam} I.~A.,   {Roberts} S.~J.,  2018,
  \mn@doi [\mnras] {10.1093/mnras/stx3187}, \href
  {https://ui.adsabs.harvard.edu/\#abs/2018MNRAS.475..331G} {475, 331}

\bibitem[\protect\citeauthoryear{{Hale}, {Jarvis}, {Delvecchio}, {Hatfield},
  {Novak}, {Smol{\v c}i{\'c}}  \& {Zamorani}}{{Hale} et~al.}{2018}]{hale2017}
{Hale} C.~L.,  {Jarvis} M.~J.,  {Delvecchio} I.,  {Hatfield} P.~W.,  {Novak}
  M.,  {Smol{\v c}i{\'c}} V.,   {Zamorani} G.,  2018, \mn@doi [\mnras]
  {10.1093/mnras/stx2954}, \href
  {http://adsabs.harvard.edu/abs/2018MNRAS.474.4133H} {474, 4133}

\bibitem[\protect\citeauthoryear{{Hardcastle}}{{Hardcastle}}{2004}]{Hardcastle2004}
{Hardcastle} M.~J.,  2004, \mn@doi [\aap] {10.1051/0004-6361:20035605}, \href
  {https://ui.adsabs.harvard.edu/#abs/2004A&A...414..927H} {414, 927}

\bibitem[\protect\citeauthoryear{{Hatfield}, {Bowler}, {Jarvis}  \&
  {Hale}}{{Hatfield} et~al.}{2018}]{Hatfield2018}
{Hatfield} P.~W.,  {Bowler} R.~A.~A.,  {Jarvis} M.~J.,   {Hale} C.~L.,  2018,
  \mn@doi [\mnras] {10.1093/mnras/sty856}, \href
  {https://ui.adsabs.harvard.edu/abs/2018MNRAS.477.3760H} {477, 3760}

\bibitem[\protect\citeauthoryear{{Heckman} \& {Best}}{{Heckman} \&
  {Best}}{2014}]{Heckman2014AGN}
{Heckman} T.~M.,  {Best} P.~N.,  2014, \mn@doi [Annual Review of Astronomy and
  Astrophysics] {10.1146/annurev-astro-081913-035722}, \href
  {https://ui.adsabs.harvard.edu/#abs/2014ARA&A..52..589H} {52, 589}

\bibitem[\protect\citeauthoryear{{Huterer}, {Knox}  \& {Nichol}}{{Huterer}
  et~al.}{2001}]{Huterer2001Cls}
{Huterer} D.,  {Knox} L.,   {Nichol} R.~C.,  2001, \mn@doi [\apj]
  {10.1086/323328}, \href
  {https://ui.adsabs.harvard.edu/#abs/2001ApJ...555..547H} {555, 547}

\bibitem[\protect\citeauthoryear{{Janssen}, {R{\"o}ttgering}, {Best}  \&
  {Brinchmann}}{{Janssen} et~al.}{2012}]{Janssen2012}
{Janssen} R.~M.~J.,  {R{\"o}ttgering} H.~J.~A.,  {Best} P.~N.,   {Brinchmann}
  J.,  2012, \mn@doi [\aap] {10.1051/0004-6361/201219052}, \href
  {https://ui.adsabs.harvard.edu/#abs/2012A&A...541A..62J} {541, A62}

\bibitem[\protect\citeauthoryear{{Jarvis}, {Bacon}, {Blake}, {Brown},
  {Lindsay}, {Raccanelli}, {Santos}  \& {Schwarz}}{{Jarvis}
  et~al.}{2015}]{Jarvis2015CSKA}
{Jarvis} M.,  {Bacon} D.,  {Blake} C.,  {Brown} M.,  {Lindsay} S.,
  {Raccanelli} A.,  {Santos} M.,   {Schwarz} D.~J.,  2015, Advancing
  Astrophysics with the Square Kilometre Array (AASKA14), \href
  {http://adsabs.harvard.edu/abs/2015aska.confE..18J} {p.~18}

\bibitem[\protect\citeauthoryear{{Jarvis} et~al.,}{{Jarvis}
  et~al.}{2016}]{Jarvis2016}
{Jarvis} M.,  et~al., 2016, in Proceedings of MeerKAT Science: On the Pathway
  to the SKA. 25-27 May, 2016 Stellenbosch, South Africa (MeerKAT2016).. p.~6
  (\mn@eprint {arXiv} {1709.01901})

\bibitem[\protect\citeauthoryear{{Jeong}, {Schmidt}  \& {Hirata}}{{Jeong}
  et~al.}{2012}]{Jeong2012}
{Jeong} D.,  {Schmidt} F.,   {Hirata} C.~M.,  2012, \mn@doi [\prd]
  {10.1103/PhysRevD.85.023504}, \href
  {https://ui.adsabs.harvard.edu/\#abs/2012PhRvD..85b3504J} {85, 023504}

\bibitem[\protect\citeauthoryear{{Joudaki}, {Dor{\'e}}, {Ferramacho},
  {Kaplinghat}  \& {Santos}}{{Joudaki} et~al.}{2011}]{Joudaki2011IM}
{Joudaki} S.,  {Dor{\'e}} O.,  {Ferramacho} L.,  {Kaplinghat} M.,   {Santos}
  M.~G.,  2011, \mn@doi [Physical Review Letters]
  {10.1103/PhysRevLett.107.131304}, \href
  {http://adsabs.harvard.edu/abs/2011PhRvL.107m1304J} {107, 131304}

\bibitem[\protect\citeauthoryear{{Karagiannis}, {Lazanu}, {Liguori},
  {Raccanelli}, {Bartolo}  \& {Verde}}{{Karagiannis}
  et~al.}{2018}]{Karagiannis2018Bispec}
{Karagiannis} D.,  {Lazanu} A.,  {Liguori} M.,  {Raccanelli} A.,  {Bartolo} N.,
    {Verde} L.,  2018, \mn@doi [\mnras] {10.1093/mnras/sty1029}, \href
  {http://adsabs.harvard.edu/abs/2018MNRAS.478.1341K} {478, 1341}

\bibitem[\protect\citeauthoryear{{Komatsu} et~al.,}{{Komatsu}
  et~al.}{2003}]{Komatsu2003WMAP}
{Komatsu} E.,  et~al., 2003, \mn@doi [\apjs] {10.1086/377220}, \href
  {http://adsabs.harvard.edu/abs/2003ApJS..148..119K} {148, 119}

\bibitem[\protect\citeauthoryear{{LSST Science Collaboration} et~al.,}{{LSST
  Science Collaboration} et~al.}{2009}]{lsstbook}
{LSST Science Collaboration} et~al., 2009, preprint, \href
  {http://adsabs.harvard.edu/abs/2009arXiv0912.0201L} {} (\mn@eprint {arXiv}
  {0912.0201})

\bibitem[\protect\citeauthoryear{{Leistedt}, {Peiris}  \& {Roth}}{{Leistedt}
  et~al.}{2014}]{Leistedt2014}
{Leistedt} B.,  {Peiris} H.~V.,   {Roth} N.,  2014, \mn@doi [\prl]
  {10.1103/PhysRevLett.113.221301}, \href
  {https://ui.adsabs.harvard.edu/abs/2014PhRvL.113v1301L} {113, 221301}

\bibitem[\protect\citeauthoryear{{Li} \& {Ma}}{{Li} \& {Ma}}{2017}]{Li2017IM}
{Li} Y.-C.,  {Ma} Y.-Z.,  2017, \mn@doi [\prd] {10.1103/PhysRevD.96.063525},
  \href {http://adsabs.harvard.edu/abs/2017PhRvD..96f3525L} {96, 063525}

\bibitem[\protect\citeauthoryear{{Lima}, {Cunha}, {Oyaizu}, {Frieman}, {Lin}
  \& {Sheldon}}{{Lima} et~al.}{2008}]{Lima2008Nz}
{Lima} M.,  {Cunha} C.~E.,  {Oyaizu} H.,  {Frieman} J.,  {Lin} H.,   {Sheldon}
  E.~S.,  2008, \mn@doi [\mnras] {10.1111/j.1365-2966.2008.13510.x}, \href
  {http://adsabs.harvard.edu/abs/2008MNRAS.390..118L} {390, 118}

\bibitem[\protect\citeauthoryear{{Lindsay} et~al.,}{{Lindsay}
  et~al.}{2014}]{Lindsay2014}
{Lindsay} S.~N.,  et~al., 2014, \mn@doi [\mnras] {10.1093/mnras/stu354}, \href
  {https://ui.adsabs.harvard.edu/#abs/2014MNRAS.440.1527L} {440, 1527}

\bibitem[\protect\citeauthoryear{{Lukic}, {Br{\"u}ggen}, {Banfield}, {Wong},
  {Rudnick}, {Norris}  \& {Simmons}}{{Lukic} et~al.}{2018}]{Lukic2018}
{Lukic} V.,  {Br{\"u}ggen} M.,  {Banfield} J.~K.,  {Wong} O.~I.,  {Rudnick} L.,
   {Norris} R.~P.,   {Simmons} B.,  2018, \mn@doi [\mnras]
  {10.1093/mnras/sty163}, \href
  {https://ui.adsabs.harvard.edu/abs/2018MNRAS.476..246L} {476, 246}

\bibitem[\protect\citeauthoryear{{Lyth}, {Ungarelli}  \& {Wands}}{{Lyth}
  et~al.}{2003}]{Lyth2003}
{Lyth} D.~H.,  {Ungarelli} C.,   {Wands} D.,  2003, \mn@doi [\prd]
  {10.1103/PhysRevD.67.023503}, \href
  {https://ui.adsabs.harvard.edu/\#abs/2003PhRvD..67b3503L} {67, 023503}

\bibitem[\protect\citeauthoryear{{Magliocchetti}, {Popesso}, {Brusa},
  {Salvato}, {Laigle}, {McCracken}  \& {Ilbert}}{{Magliocchetti}
  et~al.}{2017}]{Magliocchetti2017clustering}
{Magliocchetti} M.,  {Popesso} P.,  {Brusa} M.,  {Salvato} M.,  {Laigle} C.,
  {McCracken} H.~J.,   {Ilbert} O.,  2017, \mn@doi [\mnras]
  {10.1093/mnras/stw2541}, \href
  {http://adsabs.harvard.edu/abs/2017MNRAS.464.3271M} {464, 3271}

\bibitem[\protect\citeauthoryear{{Maldacena}}{{Maldacena}}{2003}]{Maldacena2003}
{Maldacena} J.,  2003, \mn@doi [Journal of High Energy Physics]
  {10.1088/1126-6708/2003/05/013}, \href
  {https://ui.adsabs.harvard.edu/\#abs/2003JHEP...05..013M} {2003, 013}

\bibitem[\protect\citeauthoryear{{Matarrese} \& {Verde}}{{Matarrese} \&
  {Verde}}{2008}]{Matarrese2008LSS}
{Matarrese} S.,  {Verde} L.,  2008, \mn@doi [\apjl] {10.1086/587840}, \href
  {http://adsabs.harvard.edu/abs/2008ApJ...677L..77M} {677, L77}

\bibitem[\protect\citeauthoryear{{Matthews} \& {Newman}}{{Matthews} \&
  {Newman}}{2010}]{Matthews2010}
{Matthews} D.~J.,  {Newman} J.~A.,  2010, \mn@doi [\apj]
  {10.1088/0004-637X/721/1/456}, \href
  {https://ui.adsabs.harvard.edu/\#abs/2010ApJ...721..456M} {721, 456}

\bibitem[\protect\citeauthoryear{{M{\'e}nard}, {Scranton}, {Schmidt},
  {Morrison}, {Jeong}, {Budavari}  \& {Rahman}}{{M{\'e}nard}
  et~al.}{2013}]{Menard2013}
{M{\'e}nard} B.,  {Scranton} R.,  {Schmidt} S.,  {Morrison} C.,  {Jeong} D.,
  {Budavari} T.,   {Rahman} M.,  2013, arXiv e-prints, \href
  {https://ui.adsabs.harvard.edu/\#abs/2013arXiv1303.4722M} {p.
  arXiv:1303.4722}

\bibitem[\protect\citeauthoryear{{Mueller}, {Percival}  \& {Ruggeri}}{{Mueller}
  et~al.}{2017}]{Mueller2017Opt}
{Mueller} E.-M.,  {Percival} W.~J.,   {Ruggeri} R.,  2017, preprint, \href
  {http://adsabs.harvard.edu/abs/2017arXiv170205088M} {} (\mn@eprint {arXiv}
  {1702.05088})

\bibitem[\protect\citeauthoryear{{Murray}, {Power}  \& {Robotham}}{{Murray}
  et~al.}{2013}]{Murray2013HMFcalc}
{Murray} S.~G.,  {Power} C.,   {Robotham} A.~S.~G.,  2013, \mn@doi [Astronomy
  and Computing] {10.1016/j.ascom.2013.11.001}, \href
  {http://adsabs.harvard.edu/abs/2013A%26C.....3...23M} {3, 23}

\bibitem[\protect\citeauthoryear{{Newman}}{{Newman}}{2008}]{Newman2008}
{Newman} J.~A.,  2008, \mn@doi [\apj] {10.1086/589982}, \href
  {https://ui.adsabs.harvard.edu/\#abs/2008ApJ...684...88N} {684, 88}

\bibitem[\protect\citeauthoryear{{Planck Collaboration} et~al.,}{{Planck
  Collaboration} et~al.}{2016}]{Planck2016PNG}
{Planck Collaboration} et~al., 2016, \mn@doi [\aap]
  {10.1051/0004-6361/201525836}, \href
  {http://adsabs.harvard.edu/abs/2016A%26A...594A..17P} {594, A17}

\bibitem[\protect\citeauthoryear{{Planck Collaboration} et~al.,}{{Planck
  Collaboration} et~al.}{2019}]{Planck2018PNG}
{Planck Collaboration} et~al., 2019, arXiv e-prints, \href
  {https://ui.adsabs.harvard.edu/abs/2019arXiv190505697P} {p. arXiv:1905.05697}

\bibitem[\protect\citeauthoryear{{Quinn}, {Axelrod}, {Bird}, {Dodson}, {Szalay}
   \& {Wicenec}}{{Quinn} et~al.}{2015}]{Quinn2015}
{Quinn} P.,  {Axelrod} T.,  {Bird} I.,  {Dodson} R.,  {Szalay} A.,   {Wicenec}
  A.,  2015, in Advancing Astrophysics with the Square Kilometre Array
  (AASKA14). p.~147 (\mn@eprint {arXiv} {1501.05367})

\bibitem[\protect\citeauthoryear{{Ramos Almeida}, {Bessiere}, {Tadhunter},
  {Inskip}, {Morganti}, {Dicken}, {Gonz{\'a}lez-Serrano}  \& {Holt}}{{Ramos
  Almeida} et~al.}{2013}]{RamosAlmeida2013}
{Ramos Almeida} C.,  {Bessiere} P.~S.,  {Tadhunter} C.~N.,  {Inskip} K.~J.,
  {Morganti} R.,  {Dicken} D.,  {Gonz{\'a}lez-Serrano} J.~I.,   {Holt} J.,
  2013, \mn@doi [\mnras] {10.1093/mnras/stt1595}, \href
  {https://ui.adsabs.harvard.edu/#abs/2013MNRAS.436..997R} {436, 997}

\bibitem[\protect\citeauthoryear{{Ross} et~al.,}{{Ross}
  et~al.}{2013}]{Ross2013BOSS}
{Ross} A.~J.,  et~al., 2013, \mn@doi [\mnras] {10.1093/mnras/sts094}, \href
  {http://adsabs.harvard.edu/abs/2013MNRAS.428.1116R} {428, 1116}

\bibitem[\protect\citeauthoryear{{S{\'a}nchez} \& {Bernstein}}{{S{\'a}nchez} \&
  {Bernstein}}{2019}]{Sanchez2019}
{S{\'a}nchez} C.,  {Bernstein} G.~M.,  2019, \mn@doi [\mnras]
  {10.1093/mnras/sty3222}, \href
  {https://ui.adsabs.harvard.edu/\#abs/2019MNRAS.483.2801S} {483, 2801}

\bibitem[\protect\citeauthoryear{{Santos} et~al.,}{{Santos}
  et~al.}{2015}]{Santos2015}
{Santos} M.,  et~al., 2015, in Advancing Astrophysics with the Square Kilometre
  Array (AASKA14). p.~19 (\mn@eprint {arXiv} {1501.03989})

\bibitem[\protect\citeauthoryear{{Schmidt}, {M{\'e}nard}, {Scranton},
  {Morrison}  \& {McBride}}{{Schmidt} et~al.}{2013}]{Schmidt2013}
{Schmidt} S.~J.,  {M{\'e}nard} B.,  {Scranton} R.,  {Morrison} C.,   {McBride}
  C.~K.,  2013, \mn@doi [\mnras] {10.1093/mnras/stt410}, \href
  {https://ui.adsabs.harvard.edu/\#abs/2013MNRAS.431.3307S} {431, 3307}

\bibitem[\protect\citeauthoryear{{Seljak}}{{Seljak}}{2009}]{Seljak2009MT}
{Seljak} U.,  2009, \mn@doi [Physical Review Letters]
  {10.1103/PhysRevLett.102.021302}, \href
  {http://adsabs.harvard.edu/abs/2009PhRvL.102b1302S} {102, 021302}

\bibitem[\protect\citeauthoryear{{Sheth} \& {Tormen}}{{Sheth} \&
  {Tormen}}{1999}]{Sheth1999pq}
{Sheth} R.~K.,  {Tormen} G.,  1999, \mn@doi [\mnras]
  {10.1046/j.1365-8711.1999.02692.x}, \href
  {https://ui.adsabs.harvard.edu/#abs/1999MNRAS.308..119S} {308, 119}

\bibitem[\protect\citeauthoryear{{Slosar}, {Hirata}, {Seljak}, {Ho}  \&
  {Padmanabhan}}{{Slosar} et~al.}{2008}]{Slosar2008PNG}
{Slosar} A.,  {Hirata} C.,  {Seljak} U.,  {Ho} S.,   {Padmanabhan} N.,  2008,
  \mn@doi [\jcap] {10.1088/1475-7516/2008/08/031}, \href
  {http://adsabs.harvard.edu/abs/2008JCAP...08..031S} {8, 031}

\bibitem[\protect\citeauthoryear{{Smol{\v{c}}i{\'c}}
  et~al.,}{{Smol{\v{c}}i{\'c}} et~al.}{2017a}]{Smolcic2017VLACOSMOSb}
{Smol{\v{c}}i{\'c}} V.,  et~al., 2017a, \mn@doi [\aap]
  {10.1051/0004-6361/201630223}, \href
  {https://ui.adsabs.harvard.edu/#abs/2017A&A...602A...2S} {602, A2}

\bibitem[\protect\citeauthoryear{{Smol{\v{c}}i{\'c}}
  et~al.,}{{Smol{\v{c}}i{\'c}} et~al.}{2017b}]{Smolcic2017VLACOSMOSa}
{Smol{\v{c}}i{\'c}} V.,  et~al., 2017b, \mn@doi [\aap]
  {10.1051/0004-6361/201730685}, \href
  {https://ui.adsabs.harvard.edu/#abs/2017A&A...602A...6S} {602, A6}

\bibitem[\protect\citeauthoryear{{Soo} et~al.,}{{Soo} et~al.}{2018}]{Soo2018}
{Soo} J. Y.~H.,  et~al., 2018, \mn@doi [\mnras] {10.1093/mnras/stx3201}, \href
  {https://ui.adsabs.harvard.edu/\#abs/2018MNRAS.475.3613S} {475, 3613}

\bibitem[\protect\citeauthoryear{{Square Kilometre Array Cosmology Science
  Working Group} et~al.,}{{Square Kilometre Array Cosmology Science Working
  Group} et~al.}{2018}]{SKARedBook2018}
{Square Kilometre Array Cosmology Science Working Group} et~al., 2018, arXiv
  e-prints, \href {https://ui.adsabs.harvard.edu/abs/2018arXiv181102743S} {p.
  arXiv:1811.02743}

\bibitem[\protect\citeauthoryear{{Tasse}, {Best}, {R{\"o}ttgering}  \& {Le
  Borgne}}{{Tasse} et~al.}{2008}]{Tasse2008}
{Tasse} C.,  {Best} P.~N.,  {R{\"o}ttgering} H.,   {Le Borgne} D.,  2008,
  \mn@doi [\aap] {10.1051/0004-6361:20079299}, \href
  {https://ui.adsabs.harvard.edu/#abs/2008A&A...490..893T} {490, 893}

\bibitem[\protect\citeauthoryear{{Tegmark}, {Taylor}  \& {Heavens}}{{Tegmark}
  et~al.}{1997}]{Tegmark1997Fisher}
{Tegmark} M.,  {Taylor} A.~N.,   {Heavens} A.~F.,  1997, \mn@doi [\apj]
  {10.1086/303939}, \href {http://adsabs.harvard.edu/abs/1997ApJ...480...22T}
  {480, 22}

\bibitem[\protect\citeauthoryear{{Tellarini}, {Ross}, {Tasinato}  \&
  {Wands}}{{Tellarini} et~al.}{2016}]{Tellarini2016PNG}
{Tellarini} M.,  {Ross} A.~J.,  {Tasinato} G.,   {Wands} D.,  2016, \mn@doi
  [\jcap] {10.1088/1475-7516/2016/06/014}, \href
  {http://adsabs.harvard.edu/abs/2016JCAP...06..014T} {6, 014}

\bibitem[\protect\citeauthoryear{{The Dark Energy Survey Collaboration}}{{The
  Dark Energy Survey Collaboration}}{2005}]{dark2005dark}
{The Dark Energy Survey Collaboration} 2005, ArXiv Astrophysics e-prints, \href
  {http://adsabs.harvard.edu/abs/2005astro.ph.10346T} {}

\bibitem[\protect\citeauthoryear{{Uhlemann}, {Pajer}, {Pichon}, {Nishimichi},
  {Codis}  \& {Bernardeau}}{{Uhlemann} et~al.}{2018}]{Uhlemann2018sphere}
{Uhlemann} C.,  {Pajer} E.,  {Pichon} C.,  {Nishimichi} T.,  {Codis} S.,
  {Bernardeau} F.,  2018, \mn@doi [\mnras] {10.1093/mnras/stx2623}, \href
  {http://adsabs.harvard.edu/abs/2018MNRAS.474.2853U} {474, 2853}

\bibitem[\protect\citeauthoryear{{Urry} \& {Padovani}}{{Urry} \&
  {Padovani}}{1995}]{Urry1995}
{Urry} C.~M.,  {Padovani} P.,  1995, \mn@doi [\pasp] {10.1086/133630}, \href
  {http://adsabs.harvard.edu/abs/1995PASP..107..803U} {107, 803}

\bibitem[\protect\citeauthoryear{{Welling}, {van der Woude}  \&
  {Pajer}}{{Welling} et~al.}{2016}]{Welling2016noise}
{Welling} Y.,  {van der Woude} D.,   {Pajer} E.,  2016, \mn@doi [\jcap]
  {10.1088/1475-7516/2016/08/044}, \href
  {http://adsabs.harvard.edu/abs/2016JCAP...08..044W} {8, 044}

\bibitem[\protect\citeauthoryear{{Wilman} et~al.,}{{Wilman}
  et~al.}{2008}]{Wilman2008SKADS}
{Wilman} R.~J.,  et~al., 2008, \mn@doi [\mnras]
  {10.1111/j.1365-2966.2008.13486.x}, \href
  {https://ui.adsabs.harvard.edu/#abs/2008MNRAS.388.1335W} {388, 1335}

\bibitem[\protect\citeauthoryear{{Witzemann}, {Alonso}, {Fonseca}  \&
  {Santos}}{{Witzemann} et~al.}{2018}]{Witzemann2018}
{Witzemann} A.,  {Alonso} D.,  {Fonseca} J.,   {Santos} M.~G.,  2018, arXiv
  e-prints, \href {https://ui.adsabs.harvard.edu/\#abs/2018arXiv180803093W} {p.
  arXiv:1808.03093}

\bibitem[\protect\citeauthoryear{{Xia}, {Viel}, {Baccigalupi}, {De Zotti},
  {Matarrese}  \& {Verde}}{{Xia} et~al.}{2010}]{Xia2010NVSS}
{Xia} J.-Q.,  {Viel} M.,  {Baccigalupi} C.,  {De Zotti} G.,  {Matarrese} S.,
  {Verde} L.,  2010, \mn@doi [\apjl] {10.1088/2041-8205/717/1/L17}, \href
  {http://adsabs.harvard.edu/abs/2010ApJ...717L..17X} {717, L17}

\bibitem[\protect\citeauthoryear{{Yamauchi}, {Takahashi}  \&
  {Oguri}}{{Yamauchi} et~al.}{2014}]{Yamauchi2014MT}
{Yamauchi} D.,  {Takahashi} K.,   {Oguri} M.,  2014, \mn@doi [\prd]
  {10.1103/PhysRevD.90.083520}, \href
  {http://adsabs.harvard.edu/abs/2014PhRvD..90h3520Y} {90, 083520}

\bibitem[\protect\citeauthoryear{{Yoo}}{{Yoo}}{2010}]{Yoo2010}
{Yoo} J.,  2010, \mn@doi [\prd] {10.1103/PhysRevD.82.083508}, \href
  {https://ui.adsabs.harvard.edu/\#abs/2010PhRvD..82h3508Y} {82, 083508}

\bibitem[\protect\citeauthoryear{Zaldarriaga}{Zaldarriaga}{2004}]{Zaldarriaga2004}
Zaldarriaga M.,  2004, \mn@doi [Phys. Rev. D] {10.1103/PhysRevD.69.043508}, 69,
  043508

\bibitem[\protect\citeauthoryear{{de Putter} \& {Dor{\'e}}}{{de Putter} \&
  {Dor{\'e}}}{2017}]{dePutter2017}
{de Putter} R.,  {Dor{\'e}} O.,  2017, \mn@doi [\prd]
  {10.1103/PhysRevD.95.123513}, \href
  {https://ui.adsabs.harvard.edu/\#abs/2017PhRvD..95l3513D} {95, 123513}

\bibitem[\protect\citeauthoryear{{de Putter}, {Gleyzes}  \& {Dor{\'e}}}{{de
  Putter} et~al.}{2017}]{dePutter2017b}
{de Putter} R.,  {Gleyzes} J.,   {Dor{\'e}} O.,  2017, \mn@doi [\prd]
  {10.1103/PhysRevD.95.123507}, \href
  {https://ui.adsabs.harvard.edu/abs/2017PhRvD..95l3507D} {95, 123507}

\makeatother
\end{thebibliography}

\bsp	
\label{lastpage}
\end{document}